# Reclaiming human machine nature

Didier FASS

ICN Business School
MOSEL LORIA UNIVERSITÉ de Lorraine
Campus scientifique BP239
Vandoeuvre-lès-Nancy Cedex
France
`Didier.fass@loria.fr`

**Abstract.** Extending and modifying his domain of life by artifact production is one of the main characteristics of humankind. From the first hominid, who used a wood stick or a stone for extending his upper limbs and augmenting his gesture strength, to current systems engineers who used technologies for augmenting human cognition, perception and action, extending human body capabilities remains a big issue. From more than fifty years cybernetics, computer and cognitive sciences have imposed only one reductionist model of human machine systems: cognitive systems. Inspired by philosophy, behaviorist psychology and the information treatment metaphor, the cognitive system paradigm requires a function view and a functional analysis in human systems design process. According that design approach, human have been reduced to his metaphysical and functional properties in a new dualism. Human body requirements have been left to physical ergonomics or "physiology". With multidisciplinary convergence, the issues of "human-machine" systems and "human artifacts" evolve. The loss of biological and social boundaries between human organisms and interactive and informational physical artifact questions the current engineering methods and ergonomic design of cognitive systems. New developpment of human machine systems for intensive care, human space activities or bio-engineering sytems requires grounding human systems design on a renewed epistemological framework for future human systems model and evidence based "bio-engineering". In that context, reclaiming human factors, *augmented human* and human machine nature is a necessity

## 1   Introduction: Is there "any body" inside?

*Was there a cave in Platoon's head?*

The theme of the human machine is currently a topic of research and development on the one hand and techno-philosophical-anthropological other. On the one hand reductionism assumes reduce the human body to a thinking machine and its physical and computational properties, in affiliation with Descartes and de La Métrie. On the other hand, invokes a humanist ideal, a conception of human metaphysical and transcendent, from the Renaissance and Vitruvius.

Between rationalist reductionism and metaphysical and theological idealism, the theme of "human machine" and even more "augmented human" is controversial.

Between the ideal philosophical or theological, Human with a capital H, and general scientific rational realization of the abstract category of a human biological system, there is life, the body, multidimensional integrated reality and death.

How then theoretically conceive human and scientific principles of the human machine design? What we might call *augmented human* bioengineering.

### 1.1 Human machine and intensive care

Before going further, we confront a moment the reality of the medical intensive care unit. Here the reality of human-machine is a vital necessity. The human-machine makes sense for the patient and his family. It is a matter of survival or death, for which medical teams not without practical and ethical issues. Regardless of the sophistication of these machines - mechanical ventilation, haemodialysis, cardiopulmonary bypass ... a benefit / risk assessment is always necessary and difficult. There are always risks iatrogenic real, although many automatic feedback loops have been developed [6].

### 1.2 Human machine and human space activities

If you wish to send a human in the water or in space and make him or her active, the problem is different. It is no longer survival but to expand the field of life and activity of the person, the requirements of life support and domain-specific activity. Artefacts, transport modules and living arrangements and clothing (suits) must be designed to maintain:
  i. Bodily integrity and basic physiological functioning;
  ii. Relational - sensorimotor and cognitive, and operational capacity of the operator situations;
  iii. Health of the operator in a consistent functional area with the return to earth by avoiding, for example, the risk of embolism in the plunger or cardiovascular collapse with the astronaut.

### 1.3 Human machine and convergence

With current or interactive cognitive systems [8] [12] of the smartphone to the cockpit of an airplane and the resuscitation room, or with technical systems to be operated or remotely operated by a human, the boundaries between the human and artifice, produced by human, fade. With the convergence multidisciplinary nano-bio-info-cogno (NBIC) [13] [14] amplifies this dynamic. Future implantable nano-biotechnological systems, wearable technologies and ambient intelligence and ubiquitous systems for civilian applications at all stages of life or defence let imagine new benefits and new risks to master knowledge.

This disappearance of boundaries between human biological and social and interactive and informational physical artifact questions the current engineering methods and ergonomic design of cognitive systems [1] [10] and their scientific basis.

Current and future developments in human assisted supplemented, repaired, expanded or increased not only pose new scientific and technical challenges but also new epistemological questions.

### 1.4 Reclaiming human machine nature

Currently, the human "biopsychosocial", in the words of Henri Laborit [11] is reduced in a reciprocal and symmetrical metaphorical relationship to cognitive artifacts-computo-logical-symbolic being disembodied. Although this reductionist design of human-system remains dualistic. The philosophical question of duality "mind body" has been replaced by the question of duality "body brain", where the brain is designed as a cognitive computational machine or independent of its organic substrate that processes information with loops feedback [7]. This vision is inspired by Turing and von Neumann's machines, by Shannon and Weaver and Wiener' theory of communication, and by automatic and cybernetics [16].

With multidisciplinary convergence, issues of "human-machine" systems and "human artifacts" evolve. Behavioral and cognitive heuristics metaphors, if they remain productive for the design of automatic systems generate new questions and problems that tell us to question the theoretical and experimental conceptual foundations of the "human machine". The correct design and safe technology of techniques and artefacts of "augmented human", require a system of knowledge and description revisited, without (much) of ideological and metaphorical a priori. It is for us to develop a framework for integrating artefacts with human as a matter of coupling in the multi-scale dimensions of two systems of different nature: human, biological and anthropological, and the artefact, physical and logical-symbolic.

Understanding this synthetic hybridization requires a new conceptual apparatus and a new knowledge system of the human systems integration (HIS), capable of thinking the human machine and new practices, the model and the test as a whole structurally and functionally integrated: *an epistemology of extending the area of life and human machine nature.*

## 2 Functional parameters of artificial system design: reclaiming human factors

The dominant paradigm of design and description of the human machine systems are generally outcome of behavioral and cognitive approach. They are based on a functional approach. The human is reduced, in a reciprocal and symmetrical artifacts metaphorical relation, to a disembodied computo-logical-symbolic "cognitive-being". According cognitive ergonomics, requierement engineering for human in-the-loop artificial and automation systems design is reduced to its cognitive functions.

Knowing, reasoning, understanding, planning, deciding, problem solving, analyzing, synthesizing, monitoring, assessing, checking, verifying, judging… are some the instantiation of cognitive function assumption. They refer to an agent's capacity to process or compute thoughts. According agency philosophy and artificial intelligence, an agent is an entity capable of perception, information processing or computing, and action and whose individual or collective activity is goal oriented and adapted to an environment.

This utilitarian approach is related to functionalist concepts of cognitive function, cognitive systems and joint cognitive systems [10]. Even if its wants this reductionist conception human-systems remains dualistic. This vision is inspired by Turing [Turing1936] and von Neumann's machine theory, Shannon and Weaver and Wiener's theories of communication, automatic and cybernetics

Methods and current tools of systems engineering, in particular "systems of systems" are derived from cybernetics, science and computer technology and cognition, and human factors with regard to the "systems man-in-the-loop". To do this, they represent the system through technical and managerial components approach in describing the relationship and interaction through physical interfaces and communications devices. Interaction is seen as a process of communication between components of the system reduced to each other in loops "entry, data processing, response". If it has demonstrated its heuristic nature, this metaphorical and reductionist approach is not sufficient to model embedded systems as a functional whole and scale relativity of space and time. Understanding and description of the organization of these organic or complex socio-technical systems require a unit of knowledge representation and modelling renewed.

To do this we propose a functional analysis of human factors framework in terms of existing built-in functions in both a psychological and physiological perspective: perception, decision, action, control and emotions (PDAC+E).

## 3    The "human factor"[1] Always Rings Twice.

Here is a proposal for a renewed operational human factors (HFs) model, based on functional parameters: Perception, Decision-making, Action and Control (PDAC) functions, which allow assessing each human agent or actor behavior.

For example, if a perception problem is identified as a cause for a poor decision, then we can question the data input canal.

The analysis needs to be related to the context, which is an operational context (traffic situation, weather) and role played by the agent.

---

[1] This is a wordplay. In french « facteur » means both factor or postman.

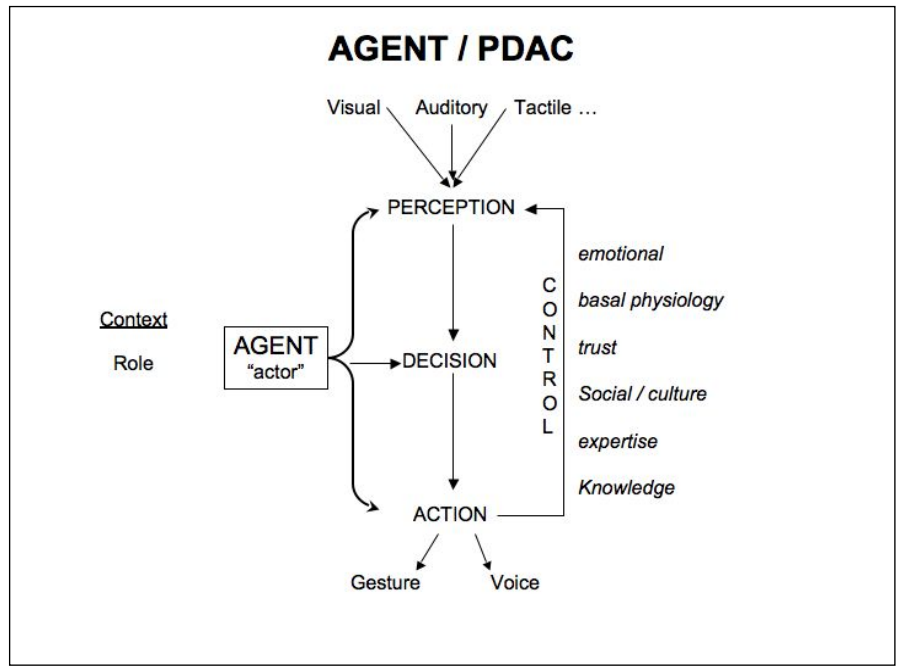

**Fig. 1.** Operational *HF model.*

An agent or an actor, involve in an activity with appropriate skills and competences and duly authorized, "plays" a role in an operational context.

The main classes of agent's function are perception, decision-making, action and control (fig. 1):

- Perception is based on visual, auditory or haptic (prioprio-tactilo-kinestetic) information.
- Decision-Making is based on reasoning and emotion. It integrates elements of perception and knowledge to answer the question: to do or not to do, what to do?
- Action is the result of an integrated and situated cognitive function. The main modes of action of an actor are gesture and voice. Interfaces and communication channels or networks mediate interactions with other agents or machines.
- Control guide thinking and behaviour in accordance with internally generated goals or plans coupled to stimuli and meaningful elements of the environment (space of actions and space of navigation) through perceptive and motor loops. Control is closely allied to attention and vigilance, basal physiology (stress, fatigue…). It also depends on knowledge, expertise and trust in collaborative activities. Emotional states influence control, also anthropological and social factors.

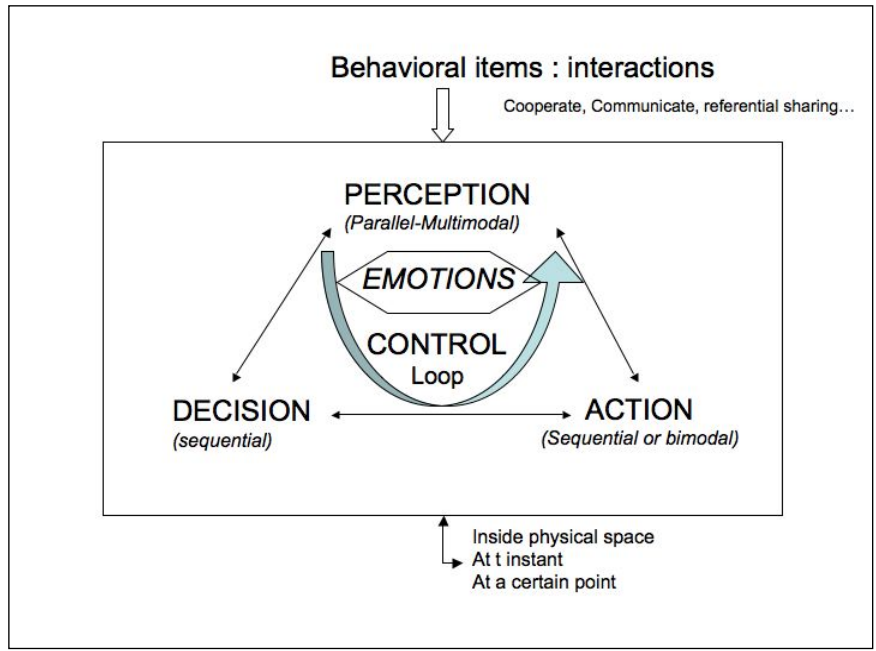

**Fig. 2.** Behavioral model: based on PDAC loop and modulated by emotions.

### 3.1 Behavioral Model

An activity is based on integration of a certain number of functions. Each human in the loop function requires several specific functional interactions for each agent for each role inside the physical space at t instant and at a certain point into the time scale relativity and space scale relativity.

Each functional interactions are related to a behavioral item or mode, such as cooperate, communicate, referential sharing…. Using this behavioral model we are able to link functions and sub functions to the observable behavior of an agent, based on the operational PDAC HF model.

Dynamics of behavioral model is based on PDAC integrated loop and modulated by emotions.

Behavioral modelling in design and simulation hold to take into account the complexity of each class of agent's fundamental function: perception is parallel and mul-timodal; decision-making is sequential, action is sequential or bimodal and control is a conscious and sequential and/or unconscious and parallel process.

Behavioral model connects PDAC-based operational HF model to collaborative and cooperative functions underlying authority sharing and distribution.

### 3.2 PDAC Analysis

What can be observed in a human in-the-loop simulation?

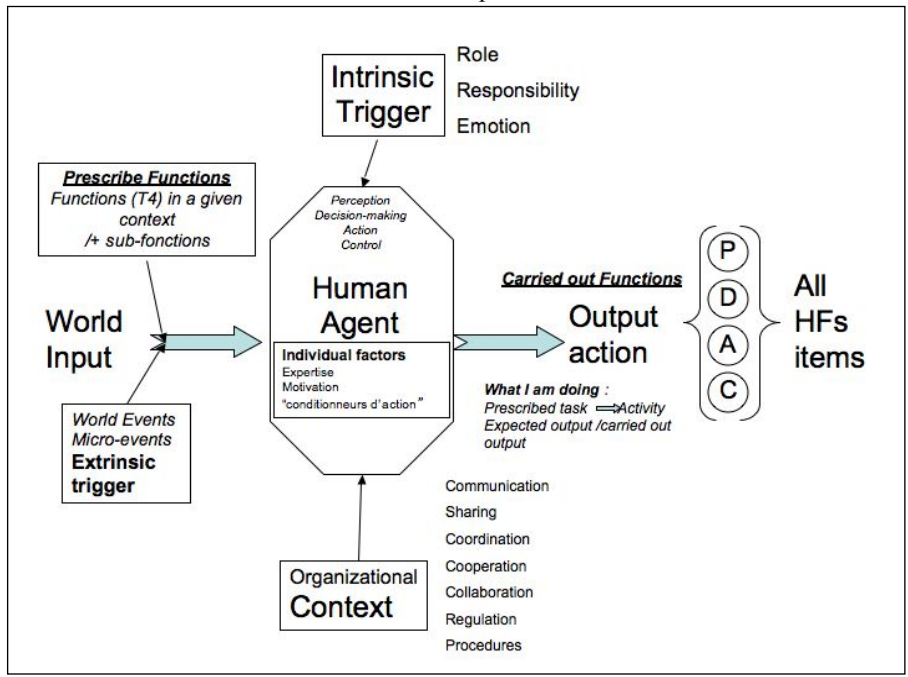

**Fig. 3.** PDAC analysis schema

From the world input, we can deduce an expected behavior related to the prescribed task. What will be observed is a result of Human activities through observable « output actions ».

> Prescribe function F:E and Carried out function
>
> $E$: world input, $S$: expected output action, $S'$: observed output action
>
> i. If $S=S'$ then $F=F'$, OK
>
> ii. If $S \neq S'$ then $F \neq F'$, then question: what HF (s) is or are involved?
>
> And what is the intrinsic trigger and organizational context (OFs)?
> Pattern variation of PDAC values allow to identify, what HFs issues are involved.

When there is a difference between expected behavior and output actions, we need to assess the HF issue that might have produced such a result: PDAC model allows to question the human basic functions and to locate the elements (intrinsic trigger and organizational context), which had an impact.

To complete that analytical methodology, we must place the PDAC analysis schema into the operational context, which must also be simulated during future experimentations.

This can be applied to a single human agent.

Now how to address collective work multi-agent environment), which is the main issue of sociotechnical systems (cooperation, authority, responsibility, and task sharing)?

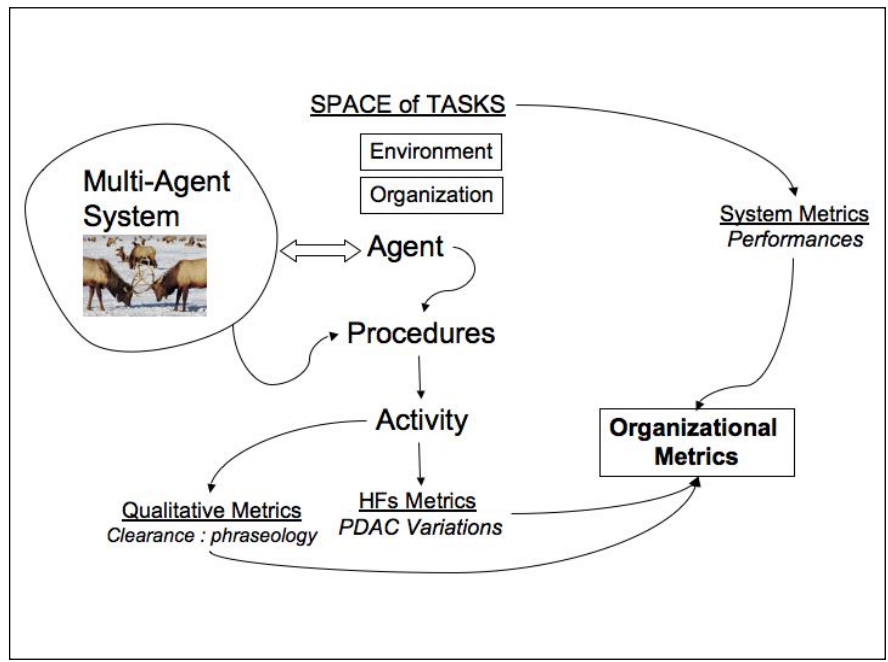

**Fig. 4.** Metrics: variation to the prescribed task

**Metrics: variation to the prescribed task.**

The situations that will be observed are multi-agent situations.

In a space of tasks, within a certain environment (both technical and organizational) the agents will « produce » their activities, based on pre-determined interaction procedures. Individual HF metrics based on PDAC model will be gathered.

At the same time it will be possible to set performance metrics (capacity, safety, efficiency) and also organizational metrics (number of vocal communications…). And also qualitative data will be extracted concerning organizational aspects.

The individual HF metrics will then be temporally and conceptually correlated with the more global measures (organizational, performance and qualitative data).

**PDAC perspective.**

To demonstrate how to use PDAC schema analysis, let's imagine a « simple task » based on very simple simulation scenario. That example presents how the PDAC analysis tools might be used.

Example: Prescribed task (Modelling, Simulation)

*"When you hear "beep" move downward the red lever {control stick}"*

- **PDAC = HFs analysis tool**: 4 elementary functions (convention)
- **Prescribed task**: prescribed function, in an organizational context that constrained relations, to an operator who is characterized by a role and a level of responsibility (legal sense)
- **World inputs**: what happen in the activity space (event or micro-event)
- **Extrinsic trigger**: *"beep"*
- **Prescribed output**
- **Fundamental model**: Agent = "PDAC",

— Aim: to explain what the agent have to do

- **Modeling prescribed function**:

— I'm waiting for **agent** hearing the "beep" (P)
— Remembering the instruction (D),
— Finding the red lever (P)
— Lever is coming downward (A)
— Agent feels the mission accomplished (C)

**Prescribed Task: PDAC analysis.**

*"When you hear "beep" move downward the red lever {control stick}"*

- Experimentation trace: « Beep is set and the red lever never come downward»
- PDAC Filter: difference Prescribed Task/ Carried out Task

What does not go in my model or what does not work in the execution of my experimentation?
    Why no P?
    Why no D?
    Why no A?
    Why no C?

However, the answer to these questions is not only analytical and functional. Another design consideration elements of man machine systems and their coupling is needed to understand how the functions and dysfunctions emerge.

## 4 Non-functional parameters of artificial system design: making sense of the organism

Current human in-the-loop and human-machine system are opened loop. Future human machine will produce regulations or counter-measures couplings generated by dynamic interfacing systems closing the loop at both organizational level and individual level. These new developments raise five major scientific and technical interdisciplinary challenges:

i. Human systems integration: from the science of systems biology and integrative physiology theory applied to human engineering systems.
ii. Epistemology and human machine systems modelling: witch system of knowledge and description?
iii. Safe design of human in-the-loop systems: numerical modelling, systems engineering and human factors.
iv. Physiological and pathophysiological modelling: what is the link between the structural elements of a system, their shapes and dynamic coupling and the emergence of functions?
v. Modelling and certification: how to validate and certificate "human machine in-the-loop" systems?

To answer these questions we need to escape the illusion of human-centred design and cognitive system reductionism.

### 4.1 Integrative epistemology and human machine design: reclaiming epistemology of coupling

Designing artificial environments and human machine systems needs to take into account both technical systems, multimodal interactions of coupling (physical, logical and informational, and biological) artificially generated and their integration into the dynamics of human behavior, cognitive, sensorimotor and emotional… and therefore in the structural and functional organization of the body: the anatomical extension of the body and the enhancement of the functions of "augmented human".

This is a problem of coupling two systems of different natures: a biological system, the human, with a physical system, the interactive artefact more or less immersive, encompassing, incorporating therefore integrative.

### 4.2 Augmented human as an hybrid organism system

The human systems integration (is to seamlessly integrate human components and passive and interactive technologies. To be safe and predictive, HIS models, the concepts of interaction and integration, methods and rules of systems engineering and design must be epistemologically well founded.

As the mathematical theory of physics and the principles underlying the mechanical or material science, e.g. technical engineering aircraft, HSI requires integration theory, a theoretical framework (conceptually and formally proven) and its principles general for coupling a biological system (experimentally proven) or with the physical artefacts (depending on the degree of complexity of the artificial system).

Some classical concept must be revised:
- Inside vs. outside
- Opened system vs. closed system
- Discrete structure, structural discontinuity
- Functional continuum, functional continuity
- Structural stability – functional viability
- Geometries of the architecture, function analysis and interactions of coupling
- System, organism

### 4.3 Biology, the human system domain

Designing augmented human [9] or human machine system as an hybrid organism, according to HSI concepts involves a paradigm shift from a metaphorical design and engineering based on usage scenarios, utility and activity - based on models and metaphysical rules of interaction and cognition, to integration engineering of dynamic structural coupling and based on an integrative theory and evidences of human machine "biology" and principles.

## 5    Conclusion – *Augmented human* bioengineering challenge

If we want to integrate human factors and their functional determinant (PDAC + E) in the design of human machine system, it should be conceived as a hybrid structural system all united by physical, logical and biological coupling interactions that generates functional continuum in space of stability of the extended anatomical body [5] and increased domain of viability of functions[3].

The logical organization of these systems is that of integrative biology and physiology of man machine systems. To conceptualize human machine systems as an organic whole organized according to the principles human systems integration grounded on theoretical systems biology requires the development of the principles and methods of a "bio-engineering" of *augmented human*.